\documentclass[floats,floatfix,amssymb,prd,twocolumn,superscriptaddress,nofootinbib,aps,10pt]{revtex4-2}
\usepackage[english]{babel}
\usepackage[utf8]{inputenc}
\usepackage{amsmath}
\usepackage{mathbbol}
\usepackage{amssymb}
\usepackage{bbold}
\usepackage{graphicx,amsfonts}
\usepackage{epsfig}
\usepackage[svgnames]{xcolor}
\definecolor{crimsonglory}{rgb}{0.75,0.0,0.2}
\usepackage[colorlinks=true,
linkcolor=crimsonglory,
urlcolor=crimsonglory,
citecolor=crimsonglory]{hyperref}
\usepackage{bm}
\usepackage{mathrsfs}
\usepackage{mathtools}
\usepackage{enumerate}
\usepackage{amsthm}
\usepackage{bbm}
\usepackage{comment}
\usepackage{physics}
\usepackage{url}
\usepackage{upgreek}
\usepackage[title]{appendix}
\usepackage{float} 
\usepackage{rotating}
\usepackage{booktabs}

\definecolor{maroon}{cmyk}{0,0.87,0.68,0.32}

\begin{document}

\title{Quasibound states of massive charged scalars around dilaton black holes in $2+1$ dimensions: Exact frequencies}

\date{\today}

\author{H. S. Vieira}
\email{horacio.santana.vieira@hotmail.com}
\email{horacio.santana-vieira@tat.uni-tuebingen.de}
\affiliation{Theoretical Astrophysics, Institute for Astronomy and Astrophysics, University of T\"{u}bingen, 72076 T\"{u}bingen, Germany}

\begin{abstract}
In this work, we investigate massive charged scalar perturbations in the background of three-dimensional dilaton black holes with a cosmological constant. We demonstrate that the wave equations governing the dynamics of these perturbations are exactly solvable, with the radial part expressible in terms of confluent Heun functions. The quasibound state frequencies are computed analytically, and we examine their dependence on the scalar field's mass and charge, as well as on the black hole's mass and electric charge. Our analysis also underscores the crucial role played by the cosmological constant in shaping the behavior of these perturbations. This specific black hole metric arises as a solution to the low-energy effective action of string theory in $2+1$ dimensions, and it holds potential for experimental realization in analog gravity systems due to the similarity between its surface gravity and that of acoustic analogs. Moreover, the analytic tractability of this system offers a valuable testing ground for exploring aspects of black hole spectroscopy, stability, and quantum field theory in curved spacetime. The exact solvability facilitates deeper insights into the interplay between geometry and matter fields in lower-dimensional gravity, where quantum gravitational effects can be more pronounced. Such studies not only enrich our understanding of dilaton gravity and its string-theoretic implications but also pave the way for potential applications in simulating black hole phenomena in laboratory settings using analog models.
\end{abstract}

\maketitle

%
%
\section{Introduction}\label{Introduction}
%
%
In black hole perturbation theory, the dynamics of scattered and/or captured waves can typically be described in three distinct stages: (i) the initial wave, which depends on the nature and characteristics of the external source causing the perturbation; (ii) the quasinormal modes and/or quasibound states, characterized by complex frequencies and referred to as quasi in contrast to normal modes or bound states, due to their intrinsically damped oscillatory nature; and (iii) a power-law tail behavior of the field, which may appear at late times depending on the spacetime and field properties \cite{Frolov:1998,LivingRevRelativity.2.2}.

In this paper, our primary interest lies in studying the spectrum of quasibound states, which are independent of the initial perturbation and depend solely on the intrinsic parameters of the black hole. Specifically, we focus on analyzing the quasibound states of massive, charged scalar fields in the background of three-dimensional charged dilaton black holes (3DCDBHs) with a cosmological constant. This investigation is motivated by the growing interest in lower-dimensional gravity models, which often provide simplified yet insightful analogs to four-dimensional black hole physics. The 3DCDBHs, in particular, offer a tractable setup for exploring the interplay between charge, mass, and scalar field interactions in curved spacetime. By studying the quasibound spectra in this context, we aim to deepen our understanding of field dynamics in lower-dimensional charged black hole spacetimes and contribute to the broader picture of black hole perturbation theory.

The positive and negative root states for a particle moving along a geodesic in a stationary background were interpreted within the framework of relativistic quantum field theory by Deruelle and Ruffini \cite{PhysLettB.52.437,LettNuovoCim.15.257}. These states are considered classical counterparts of the positive and negative energy solutions in a quantized field and are referred to as quasibound states (QBSs). In essence, QBSs are resonant modes localized between the effective potential generated by the curved spacetime geometry and spatial infinity, where they remain temporarily confined due to the potential barrier. Importantly, transitions or crossings between these states may occur, potentially satisfying the necessary conditions for particle creation in curved spacetime.

In recent years, considerable effort has been devoted to the computation and analysis of QBSs across a wide range of black hole spacetimes \cite{PhysRevD.106.024046,JHEAP.40.49,PhysRevD.109.124037,JHEAP.41.61,PhysRevD.109.084018,EurPhysJC.84.57,EurPhysJC.84.438,PhysLettB.849.138414,PhysScr.99.085023,PhysLettB.848.138373,JHEAP.43.286,EurPhysJC.84.607,PhysLettB.854.138714,EurPhysJC.84.424,EurPhysJC.84.229,EurPhysJC.84.857,JHEAP.43.132,EurPhysJC.85.352,AnnPhys.479.170051,EurPhysJC.85.270,PhysRevD.111.044055,AnnPhys.473.169898}. Parallel studies have also explored analogous behaviors in effective models of gravity, particularly in acoustic black hole systems \cite{PhysRevLett.121.061101,PhysRevA.101.022507,PhysRevD.103.045004,PhysRevD.110.044017,PhysRevB.109.035407,EurPhysJC.84.388,PhysRevD.110.104058,PhysRevD.111.104064}, where the propagation of perturbations in fluids or other media mimic aspects of black hole spacetimes \cite{PhysRevLett.46.1351,ClassQuantumGrav.15.1767,LivingRevRelativity.14.3}. Most of these investigations employed the Vieira--Bezerra--Kokkotas (VBK) method \cite{AnnPhys.373.28,PhysRevD.104.024035} to determine the QBS frequencies, although some studies have utilized purely numerical approaches.

The growing interest in QBSs stems from their crucial role in understanding various phenomena, including black hole spectroscopy, superradiance, and stability analysis. In particular, the spectrum of QBS frequencies carries valuable information about the underlying geometry and physical parameters of the spacetime. Moreover, in both astrophysical and analog gravity contexts, QBSs offer a window into processes involving horizon thermodynamics, vacuum polarization, and even the early stages of black hole evaporation. As such, ongoing research continues to refine analytical and numerical methods for studying QBSs, aiming to connect theoretical predictions with future observational data from gravitational wave detectors and laboratory analog experiments

The VBK approach aims to transform the radial part of the equation of motion into a Heun-type equation without making any prior assumptions about boundary conditions. Notably, the VBK method is not limited to Heun equations; it can also reduce the radial equation into other forms of high transcendental equations, such as the hypergeometric and Mathieu equations. Once the transformation is performed, the QBS boundary conditions are imposed on the resulting exact analytical radial solution. Specifically, these conditions require ingoing wave behavior at the (exterior) event horizon and vanishing behavior at spatial infinity.

According to the VBK formalism, these two asymptotic radial solutions are matched within their overlapping region of validity when the Heun functions reduce to polynomials. This polynomial condition is satisfied if and only if two specific truncation constraints on the power series are met. As a result, the VBK approach provides a fully analytical framework for determining the quasispectrum of black holes -- that is, for analyzing the behavior of field perturbations in both astrophysical and analog black hole systems.

In recent years, the VBK method has gained attention for its versatility in handling complex geometries and its potential applications in emerging areas such as black hole spectroscopy and gravitational wave astronomy. By offering precise analytical insights into the spectrum of perturbations, the VBK approach complements numerical methods and helps deepen our theoretical understanding of wave dynamics in curved spacetimes.

The dilaton black hole considered in the present paper was originally derived by Chan and Mann \cite{PhysRevD.50.6385} and represents a significant solution within the low-energy string-inspired Einstein--Maxwell--dilaton theory in $2+1$ dimensions, incorporating a cosmological constant. The inclusion of this cosmological constant is crucial, as it allows the black hole to feature two event horizons rather than a single one, thereby enriching its causal structure. One of the primary objectives of this study is to investigate the impact of the cosmological constant on the QBS spectra of charged dilaton black holes. The literature contains numerous analyses on scalar field perturbations in 3DCDBHs, including the quasinormal mode analysis by Fernando \cite{PhysRevD.77.124005}, investigations into particle collisions also conducted by Fernando \cite{ModPhysLettA.32.1750088}, and computations of Hawking radiation by Sakalli \cite{AstrophysSpaceSci.340.317}, among others.

Building on these foundational studies, the present work aims to provide a more comprehensive understanding of how varying the cosmological constant influences the stability and dynamical properties of the charged dilaton black hole under scalar perturbations. By examining the QBS spectra in detail, we seek to elucidate the interplay between the black hole's mass and charge, the dilaton coupling, and the cosmological constant, thereby offering new insights into the underlying physics governing these lower-dimensional gravitational systems. This analysis not only sheds light on the theoretical aspects of black hole perturbations in string-inspired models but also contributes to the broader understanding of quantum field behavior in curved spacetimes with nontrivial horizons.

The layout of this paper is as follows. In Sec.~\ref{3DCDBH}, we introduce the metric corresponding to three-dimensional charged dilaton black holes. Sec.~\ref{SWE} is devoted to solving the equation of motion for charged massive scalar fields in the specified background. In Sec.~\ref{QBSs}, we apply the VBK approach to derive the spectrum of quasibound states. Finally, Sec.~\ref{Conclusions} presents our conclusions. Throughout this work, we adopt natural units by setting $G \equiv c \equiv \hbar \equiv 1$.
%
%
\section{Three-dimensional charged dilaton black holes with cosmological constant}\label{3DCDBH}
%
%
We begin by outlining the geometry and key characteristics of the charged black hole analyzed in this work. The theoretical framework is based on the Einstein--Maxwell-dilaton theory with a cosmological constant (EMD$\Lambda$ theory), which is conformally related to the low-energy limit of string theory in $2+1$ dimensions. This theory provides a rich setting for exploring black hole solutions in lower-dimensional gravity, as discussed in what follows.

The three-dimensional Maxwell-dilaton gravity is a field theory that extends Einstein gravity in three dimensions by coupling it to both a Maxwell field (electromagnetism) and a dilaton field (a scalar field that often arises in string theory and lower-dimensional gravity models). The physical phenomena related to 3D Maxwell-dilaton gravity include: Black hole solutions, electric and magnetic charges, gravitational and gauge solitons, black hole thermodynamics, holography and Anti-de Sitter/Conformal field theory (AdS/CFT) correspondence, topological and Chern--Simons terms, dimensional reduction and string theory, supersymmetry and Bogomol'nyi--Prasad--Sommerfeld (BPS) states and hairy black holes \cite{JHEP.2012.89,PhysRevD.97.064032,JHEP.2021.58,EurPhysJC.82.600,PhysRevD.107.024043}.

The action of the EMD$\Lambda$ theory is defined by
\begin{eqnarray}
\mathcal{S}_{\rm EMD\Lambda} & = & \int d^{3}x\ \sqrt{-g}\ [R-4(\nabla\phi)^{2}-\mbox{e}^{-4\phi}F_{\rho\sigma}F^{\rho\sigma}\nonumber\\
& & +2\mbox{e}^{4\phi}\Lambda],
\label{eq:action_EMDL}
\end{eqnarray}
where $R$, $\phi$, $F_{\rho\sigma}$, and $\Lambda$ are the Ricci scalar curvature, the dilaton field parameter, the Maxwell field strength, and the cosmological constant, respectively. In this theory, black hole solutions exist only when the cosmological constant satisfies $\Lambda > 0$. Therefore, throughout the present paper, we restrict our analysis to the range $0 < \Lambda \leq 1$.

\begin{figure}[t]
	\centering
	\includegraphics[width=1\columnwidth]{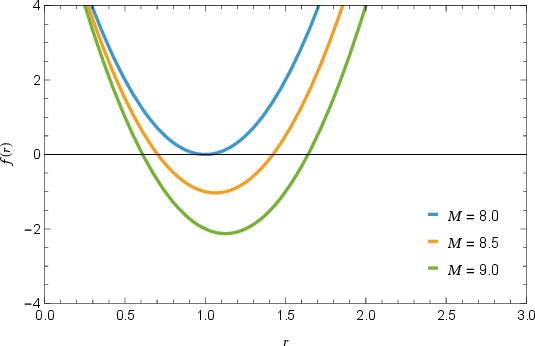}
	\caption{The metric function $f(r)$ with unitary charge and cosmological constant, varying the radial coordinate $r$ and different choices of mass $M$.}
	\label{fig:Fig1_QBSs_3DCDBH}
\end{figure}

\begin{figure}[t]
	\centering
	\includegraphics[width=1\columnwidth]{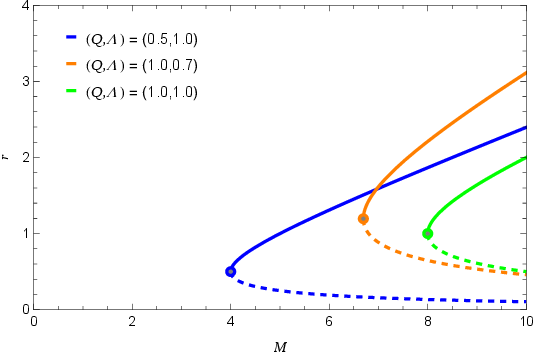}
	\caption{The event horizons $r_{\pm}$ varying the mass $M$ and different choices of the parameters $Q$ and $\Lambda$. The solid colored lines corresponds to the exterior event horizon $r_{+}$, while the dashed colored lines denote the interior event horizon $r_{-}$. The extreme values for the event horizons are marked by gray points.}
	\label{fig:Fig2_QBSs_3DCDBH}
\end{figure}

As a solution for the EMD$\Lambda$ field equations, the line element describing 3DCDBHs is given by
\begin{equation}
ds^{2} = -f(r)\ dt^{2}+\frac{4r^{2}}{f(r)}dr^{2}+r^{2}\ d\theta^{2},
\label{eq:line_element_3DCDBH}
\end{equation}
with the metric function $f(r)$ written as
\begin{equation}
f(r)=8\Lambda r^{2}-2Mr+8Q^{2},
\label{eq:metric_function_3DCDBH}
\end{equation}
where $Q$ is the charge of the black hole. In Figure \ref{fig:Fig1_QBSs_3DCDBH}, we present the behavior of the metric function for fixed values of the parameters $M$, $Q$ and $\Lambda$. We can conclude that the surface gravity equation, $f(r)=0$, has two solutions, which are given by
\begin{equation}
r_{\pm}=\frac{M \pm \sqrt{M^{2}-64Q^{2}\Lambda}}{8\Lambda}.
\label{eq:event_horizons_3DCDBH}
\end{equation}
In fact, for $M > 8Q\sqrt{\Lambda}$, the line element given by Eq.~(\ref{eq:line_element_3DCDBH}) describes a black hole, with two solutions corresponding to the exterior event horizon $r_{+}$ and the interior (Cauchy) event horizon $r_{-}$ event horizons of the 3DCDBHs. Thus, the exterior event horizon $r_{+}$ represents the outermost marginally trapped surface for outgoing photons. Figure \ref{fig:Fig2_QBSs_3DCDBH} illustrates the behavior of these horizons for fixed values of the parameters $Q$ and $\Lambda$, as the mass $M$ varies. We conclude that extreme 3DCDBHs occur when $M=8Q\sqrt{\Lambda}$; in the present paper, we focus exclusively on the non-extreme case. Additionally, this spacetime features a timelike singularity located at $r=0$; this structure plays a crucial role in determining the causal and thermodynamic properties of the black hole.

Finally, the Maxwell field strength, the dilaton field parameter, and the three-vector electromagnetic potential take, respectively, the forms
\begin{equation}
F_{rt}=\frac{Q}{r^{2}},
\label{eq:Maxwell_field_3DCDBH}
\end{equation}
\begin{equation}
\phi=-\frac{1}{2}\ln(r),
\label{eq:dilaton_field_3DCDBH}
\end{equation}
and
\begin{equation}
A_{\rho}=\frac{Q}{r}(1,0,0).
\label{eq:electromagnetic_potential_3DCDBH}
\end{equation}
Furthermore, this black hole metric is also a solution to the low energy string theory by considering the conformal transformation
\begin{equation}
g^{\rm string} = \mbox{e}^{4\phi}\ g^{\rm Einstein},
\label{eq:conformal_transformation_3DCDBH}
\end{equation}
where $g \equiv \mbox{det}(g_{\rho\sigma})$. In general, charged black hole solutions can be derived from their uncharged counterparts through the application of duality transformations; for a comprehensive review, please see Refs.~\cite{UCSBTH-92-32,JMathPhys.35.4839,GenRelativGravit.36.71}. These duality techniques serve as powerful tools in gravitational physics, allowing the extension of known solutions to a wider class of charged configurations without solving the field equations from scratch.

\subsection{Analogy between the 3D Maxwell-dilaton gravity and analog models of gravity}

The black hole solutions in 3D Maxwell-dilaton gravity do not have direct experimental realizations in the way, say, General Relativity predictions do (like gravitational waves), but they do have analog models or experimental analogs that share key physical or mathematical features. These analog models allow us to simulate aspects of black hole physics -- such as horizons, Hawking radiation, and thermodynamics -- in lab-based systems, even if the full 3D Maxwell-dilaton model is not realized. The analog models of gravity related to Maxwell-dilaton black holes can basically be divided in four classes, as follows \cite{IntJModernPhysA.18.3735,PhysRevA.78.021603,PhysRevB.86.144505,AnnPhys.327.2617,PhysRevA.88.033843}.

Acoustic black holes (or dumb holes) are fluid systems where perturbations (like sound waves) experience horizons, whose effective geometry seen by sound waves can be written in a metric form. In some setups, the metric mimics that of dilaton gravity black holes, especially in (1+1)D or (2+1)D models. The fluid's velocity field plays a role similar to the metric function. Among the experimental realizations, we can mention Bose--Einstein condensates (BECs), water tanks, and optical fibers or waveguides (for analog Hawking radiation). The key similarities are the emergent metric and the horizon behavior.

Optical analog gravity is the use of metamaterials or nonlinear optical media to mimic spacetime geometries. Light propagation in a medium with spatially varying refractive index can mimic geodesic motion in curved spacetime. Then, the dilaton field can be mimicked by spatial variation of material properties. Among the experimental realizations, we can mention the ones with nonlinear optics that has simulated event horizons and Hawking-like effects, and transformation optics that can emulate features like ergospheres and trapping surfaces.

Electrical circuit analogs are networks of inductors, capacitors, and resistors that can simulate dynamical systems including field equations. Maxwell-dilaton equations can, in principle, be mimicked by tuning circuit parameters to encode field-dependent couplings (like dilaton-modulated permittivity/permeability). They can be used to study black hole quasinormal modes or wave propagation in analog spacetimes.

In materials like graphene, low-energy excitations behave as Dirac fermions in curved space. With strain or deformation, the effective metric changes. Then, it can simulate black hole spacetimes and even simulate scalar fields via spatially varying potential. Proposals exist for simulating (2+1)D gravity-like systems, possibly including gauge and scalar field couplings.

Finally, in summary, similarities between the Maxwell-dilaton gravity and its analog model counterpart lie in the fact that event horizons from metric zeros can be simulated by sonic/optical horizons in fluids or light, scalar field modifying gravity/gauge coupling by spatially varying medium properties, Maxwell field (electric/magnetic charge)	by fluid vorticity and/or light polarization, entropy and temperature influenced by dilaton can be mimicked by analog Hawking radiation and entropy in BECs, as well as topological effects such as Chern--Simons terms and charge quantization by effective gauge fields in condensed matter. Of course, there are limitations: Analog models do not reproduce gravity itself, only certain aspects of the field equations or geometrical structure, and then they typically simulate kinematics (wave propagation) rather than dynamics (Einstein equations), and therefore, since the dilaton in 3D gravity is often tied to string theory or dimensional reduction, these deeper origins are not captured in analog systems \cite{PhysLettA.380.1,CommunPhys.3.120,ProgTheorExpPhys.2020.073E01,PhilTransRSocA.378.20190232,EurPhysJC.84.767}.

In what follows, we focus on the study of massive charged scalar perturbations in 3DCDBHs and aim to compute their QBS spectrum. To achieve this, we will employ the VBK approach, which enables us to obtain exact frequencies of QBSs. This method provides a robust framework for analyzing the stability properties and resonance structures of the black holes under scalar field perturbations. Understanding the QBS spectrum of such perturbations is crucial for probing the interaction between matter fields and the black hole background. It also sheds light on potential observational signatures and the dynamical response of these lower-dimensional black holes, which serve as valuable theoretical laboratories for exploring fundamental aspects of gravity and quantum field theory in curved spacetime.
%
%
\section{Scalar wave equation: Analytical solutions}\label{SWE}
%
%
The VBK approach is an analytical method used to investigate quasibound states of black holes, particularly those involving scalar fields or other perturbing fields around various types of black hole spacetimes. This technique is rooted in applying a Heun function-based formalism to solve the radial and angular parts of the wave equations that govern field perturbations in curved spacetime backgrounds. The key feature of this approach is its ability to handle the effective potential barriers that arise in the black hole perturbation problem, allowing for precise determination of the complex frequencies associated with quasibound states -- modes that are localized near the black hole and decay over time due to their interaction with the curved geometry. These states are of great interest in gravitational physics, especially in the context of superradiant instabilities and ultralight bosonic fields, which could have implications for new physics beyond the Standard Model.

The VBK approach is notable for its versatility in adapting to different spacetime geometries, including Schwarzschild, Kerr, and Kerr--Newman black holes, and for its effectiveness in computing frequencies even when purely numerical methods become challenging due to the complexity of the effective potential. By bridging the gap between analytical techniques and numerical accuracy, this method offers a powerful tool for exploring the spectrum of black hole perturbations in a wide array of theoretical scenarios.

We want to study some basic features of the curved spacetime geometry in Eq.~\ref{eq:line_element_3DCDBH}, in particular the propagation of massive charged scalar fields in the exterior region of this background. To perform this analysis, we start considering the covariant minimally coupled Klein--Gordon equation in the presence of an electromagnetic field, which is given by
\begin{eqnarray}
&& \biggl[\frac{1}{\sqrt{-g}}\partial_{\mu}(g^{\mu\nu}\sqrt{-g}\partial_{\nu})-ie(\partial_{\mu}A^{\mu})-2ieA^{\mu}\partial_{\mu}\nonumber\\
&& -\frac{ie}{\sqrt{-g}}A^{\mu}(\partial_{\mu}\sqrt{-g})-e^{2}A^{\mu}A_{\mu}-\mu^{2}\biggr]\Psi=0,
\label{eq:covariant_KG_equation}
\end{eqnarray}
where $\mu$ and $e$ are the mass and the charge of the scalar particle, respectively, and $\Psi=\Psi(t,r,\theta)$ is the scalar wave function. Now, by substituting Eq.~(\ref{eq:line_element_3DCDBH}) into Eq.~(\ref{eq:covariant_KG_equation}), we obtain the following partial scalar wave equation
\begin{eqnarray}
&& -\frac{1}{f(r)}\frac{\partial^{2} \Psi}{\partial t^{2}}-\frac{2ieQ}{rf(r)}\frac{\partial \Psi}{\partial t}+\frac{f(r)}{4r^{2}}\frac{\partial^{2} \Psi}{\partial r^{2}}+\frac{1}{4r^{2}}\frac{df(r)}{dr}\frac{\partial \Psi}{\partial r}\nonumber\\
&& +\frac{1}{r^{2}}\frac{\partial^{2} \Psi}{\partial \theta^{2}}+\biggl[\frac{e^{2}Q^{2}}{r^{2}f(r)}-\mu^{2}\biggr]\Psi=0.
\label{eq:scalar_wave_equation_3DCDBH}
\end{eqnarray}
By means of symmetry, we consider the following separation ansatz for the partial scalar wave function $\Psi$:
\begin{equation}
\Psi(t,r,\theta) \simeq \mbox{e}^{-i \omega t}R(r)\mbox{e}^{i m \theta},
\label{eq:ansatz_3DCDBH}
\end{equation}
where $\omega$ is the frequency (energy, in the natural units), $R(r)$ is the radial function, and $m$ is an integer running from $-\infty$ to $+\infty$, which can be called as azimuthal number, without loss of generality. Then, by substituting Eq.~(\ref{eq:ansatz_3DCDBH}) into Eq.~(\ref{eq:scalar_wave_equation_3DCDBH}), we obtain the radial master wave equation given by
\begin{eqnarray}
&& R''(r)+\biggl(\frac{1}{r-r_{+}}+\frac{1}{r-r_{-}}\biggr)R'(r)\nonumber\\
&& +\frac{4}{(r-r_{+})^{2}(r-r_{-})^{2}}[(r\omega-eQ)^{2}\nonumber\\
&& -(r-r_{+})(r-r_{-})(m^{2}+r^{2}\mu^{2})]R(r)=0,
\label{eq:radial_equation_3DCDBH}
\end{eqnarray}
where we have written the metric function as $f(r)=(r-r_{+})(r-r_{-})$. From Eq.~(\ref{eq:radial_equation_3DCDBH}), we can infer that, for superradiance to occur, the oscillation frequency of the perturbation $\omega$ must be less than a critical value at the event horizon, namely, $\omega_{\pm}=eQ/r_{\pm}$. Superradiance in charged black holes is a wave amplification phenomenon where reflected waves have a greater amplitude than incident waves. This contrasts with the usual absorption by black holes and can lead to energy extraction from the black hole; it is a wave analogue of the Penrose process for particles.

Next, we are going to apply the VBK approach to transform this radial equation (\ref{eq:radial_equation_3DCDBH}) into a confluent Heun equation \cite{Ronveaux:1995} without any boundary condition assumptions. By using the VBK approach, we essentially obtain the canonical confluent Heun equation \cite{JPhysAMathTheor.43.035203}
\begin{eqnarray}
&& y''(x)+\biggl(\alpha+\frac{1+\beta}{x}+\frac{1+\gamma}{x-1}\biggr)y'(x)\nonumber\\
&& +\biggl(\frac{\xi}{x}+\frac{\zeta}{x-1}\biggr)y(x)=0,
\label{eq:canonical_CHE}
\end{eqnarray}
with
\begin{equation}
x=\frac{r-r_{-}}{r_{+}-r_{-}},
\label{eq:x_3DCDBH}
\end{equation}
where $y(x)=\mbox{HeunC}(\alpha,\beta,\gamma,\delta,\eta;x)$ are the confluent Heun functions, with the parameters $\alpha$, $\beta$, $\gamma$, $\delta$ and $\eta$ related to $\xi$ and $\zeta$ by the following expressions
\begin{eqnarray}
\xi		& = & \frac{1}{2}(\alpha-\beta-\gamma+\alpha\beta-\beta\gamma)-\eta,\label{eq:xi_CHE}\\
\zeta	& = & \frac{1}{2}(\alpha+\beta+\gamma+\alpha\gamma+\beta\gamma)+\delta+\eta.\label{eq:zeta_CHE}
\end{eqnarray}
Here, the radial function is such that $R(x)=x^{\beta/2}(x-1)^{\gamma/2}\mbox{e}^{\alpha x/2}y(x)$, and these parameters are functions of $\omega$, $\Lambda$, $M$, $Q$ and $m$ and given by
\begin{eqnarray}
\alpha & = & -4(r_{+}-r_{-})\mu,\label{eq:alpha_3DCDBH}\\
\beta  & = & -\frac{4ir_{-}(\omega-\omega_{-})}{r_{+}-r_{-}},\label{eq:beta_3DCDBH}\\
\gamma & = & -\frac{4ir_{+}(\omega-\omega_{+})}{r_{+}-r_{-}},\label{eq:gamma_3DCDBH}\\
\delta & = & -4(r_{+}^{2}-r_{-}^{2})\mu^{2},\label{eq:delta_3DCDBH}\\
\eta   & = & -\frac{8r_{+}r_{-}(\omega-\omega_{+})(\omega-\omega_{-})}{(r_{+}-r_{-})^2}\nonumber\\
&& -4(m^{2}+r_{-}^{2}\mu^{2}).\label{eq:eta_3DCDBH}
\end{eqnarray}
For a detailed discussion on the VBK approach, we cordially invite the readers to see the Appendix A of Ref.~\cite{PhysRevD.111.104025}.
%
%
\section{Quasibound states: Exact frequencies}\label{QBSs}
%
%
There exist two linearly independent analytic solutions to the covariant minimally coupled Klein--Gordon equation near the boundaries of the geometry described by Eq.~(\ref{eq:line_element_3DCDBH}), expressed in terms of confluent Heun functions. To determine the QBS spectrum of the system, we impose two physically motivated boundary conditions on the radial solution: (i) purely ingoing waves at the exterior event horizon, and (ii) vanishing behavior at spatial infinity. By deriving the asymptotic forms of these solutions both near the event horizon and at spatial infinity, and enforcing these boundary conditions, we match the solutions within their overlapping region of validity. This matching procedure utilizes the polynomial conditions inherent to the confluent Heun functions, allowing us to extract the QBS spectrum characteristic of the background spacetime.

Furthermore, the determination of the QBS spectrum provides critical insights into the stability and resonant properties of the background geometry. These modes characterize how scalar perturbations evolve and decay over time, encoding information about the underlying gravitational field and potential well structure. Analyzing the spectral properties not only reveals the dynamical response of the system but also serves as a diagnostic tool for detecting instabilities or phase transitions in the geometry. Consequently, the study of QBSs offers a powerful probe into both classical and quantum aspects of the gravitational background under investigation.

Thus, from the first polynomial condition of the confluent Heun functions, namely, the so-called $\delta$-condition, which is given by
\begin{equation}
\frac{\delta}{\alpha}+\frac{\beta+\gamma+2}{2	}=-n,
\label{eq:delta-condition}
\end{equation}
we essentially obtain the exact QBS frequencies, i.e.,
\begin{equation}
\omega_{n}=\frac{8eQ\Lambda}{M}-i\frac{[4\Lambda(n+1)+M\mu]\sqrt{M^{2}-64Q^{2}\Lambda}}{8M\Lambda},
\label{eq:omega_3DCDBH}
\end{equation}
where $n=0,1,2,\ldots$ is the overtone number.

\begin{figure*}[t]
	\centering
	\includegraphics[scale=1]{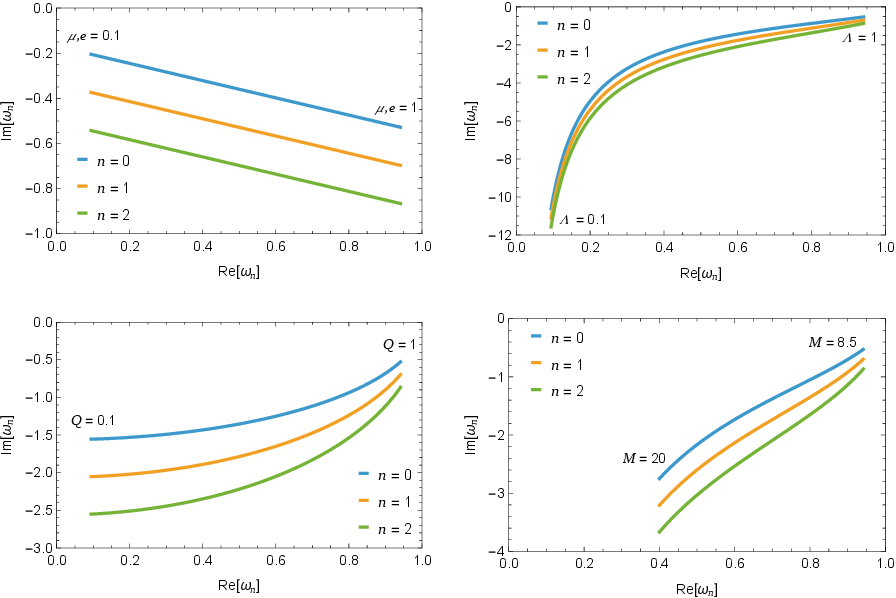}
	\caption{The first three QBS modes $n=0,1,2$. \textit{Top left panel:} The QBSs with unitary cosmological constant, black hole's mass and charge $\Lambda=M=Q=1$, varying either the scalar particle's mass $\mu=0.1,\ldots,1.0$ or charge $e=0.1,\ldots,1.0$. \textit{Top right panel:} The QBSs with unitary black hole's mass and charge $\Lambda=M=Q=1$, as well as unitary scalar particle's mass and charge $\mu=e=1$, varying the cosmological constant $\Lambda=0.1,\ldots,1.0$. \textit{Bottom left panel:} The QBSs with unitary black hole's mass, and cosmological constant $M=\Lambda=1$, as well as unitary scalar particle's mass and charge $\mu=e=1$, varying the black hole's charge $Q=0.1,\ldots,1.0$. \textit{Bottom right panel:} The QBSs with unitary black hole's charge, and cosmological constant $Q=\Lambda=1$, as well as unitary scalar particle's mass and charge $\mu=e=1$, varying the black hole's mass $M=8.5,\ldots,20$.}
	\label{fig:Fig3_QBSs_3DCDBH}
\end{figure*}

In Figure \ref{fig:Fig3_QBSs_3DCDBH}, we present, for completeness, the typical behavior of three-dimensional charged dilaton scalar QBSs. Specifically, an increase in the scalar particles mass parameter $\mu$ leads to a higher oscillation frequency (i.e., an increase in the real part of the QBS frequency), while simultaneously prolonging the perturbation's lifetime, as reflected by an increase in the magnitude of the imaginary part. Conversely, an increase in the black hole mass parameter $M$ results in a decrease of the oscillation frequency and a corresponding increase (in absolute value) of the imaginary part, indicating a faster decay rate. Furthermore, increasing both the black hole charge $Q$ and the cosmological constant $\Lambda$ causes the real part of the QBS frequencies to increase, whereas the imaginary part decreases, implying a stabilization of the perturbations within the 3DCDBH spacetime.

These behaviors reflect the intricate interplay between the black hole parameters and the scalar field properties in shaping the QBS spectrum. The sensitivity of the QBS frequencies to changes in $M$, $Q$, $\mu$, and $\Lambda$ underscores the complex dynamics governing scalar perturbations in lower-dimensional gravity models. Understanding these dependencies is crucial for probing the stability and resonant modes of charged dilaton black holes, which may shed light on their holographic duals and the nature of scalar field interactions in curved backgrounds.

This study can offer valuable insights into quantum gravity because these QBSs often exist at the intersection of quantum mechanics and gravitational physics. Studying QBSs offers a tractable way to explore systems where both gravity and quantum effects are significant. By treating these QBSs as the ``spectral fingerprints'' of quantum gravity, we can look for clues about the nature of spacetime at its most fundamental level.
%
%
\section{Final remarks}\label{Conclusions}
%
%
In this paper, we investigate the geometry of a 3DCDBH, focusing specifically on its characteristic frequencies associated with charged massive scalar QBSs, which acts like a spectroscopic fingerprint of the background spacetime, similar to how atomic spectra reveal the structure of atoms. This analysis provides valuable insights into the quantum phenomena occurring near black holes, as well as classical processes that remain difficult to observe through both astrophysical observations and analog detection methods.

Understanding these QBSs not only deepens our theoretical knowledge of black hole physics but also offers potential avenues for detecting subtle quantum effects in gravitational systems. By exploring the interplay between charge, mass, and the dilaton field in lower-dimensional models, our study paves the way for future research aimed at bridging the gap between classical general relativity and emerging quantum gravity theories.

In conclusion, Figure \ref{fig:Fig3_QBSs_3DCDBH} reveals a significant pattern in the QBS frequencies relative to all the parameters of the 3DCDBHs. Our results exhibit a complex structure that could prove valuable for future comparisons with experimental and observational data. This insight lays the groundwork for a potential analytical description of the behavior of these states, enhancing our theoretical understanding of 3DCDBHs.

Furthermore, exploring these QBS frequency patterns could provide critical clues about the underlying physical processes governing black hole perturbations and stability. By establishing a clear connection between the theoretical predictions and empirical findings, this work may contribute to the development of more refined models of black hole dynamics, shedding light on fundamental aspects of gravity and quantum field interactions in curved spacetime.

Our analysis of the QBSs naturally yields a spectrum of modes corresponding to different overtone numbers and various geometric parameters -- namely, the black hole's mass $M$ and charge $Q$, the scalar field's mass $\mu$ and charge $e$, as well as the cosmological constant $\Lambda$. Importantly, this framework imposes no a priori constraints linking these free parameters, allowing for a fully general exploration of their combined effects. The wide range of possible parameter combinations opens up a rich landscape for probing the underlying physics.

QBS lifetimes may constrain models of quantum information retention, complementing studies of Hawking radiation and black hole evaporation. Furthermore, near the exterior event horizon, the QBSs might serve as a model for soft hair, quantum memory, or even pre-Hawking radiation effects in lower-dimensional analogs of black hole evaporation.

Looking ahead, the potential for experimental and observational breakthroughs is substantial. Advances in gravitational wave astronomy, black hole imaging, and high-precision cosmological measurements can provide critical tests of these theoretical predictions. By comparing observed data with the predicted QBS spectra, researchers can deepen their understanding of black hole dynamics, shedding light on both quantum effects near the event horizon and classical gravitational phenomena. Ultimately, such interdisciplinary efforts will help bridge the gap between theory and observation, pushing the frontier of black hole physics to new horizons.
%
%
\begin{acknowledgments}
%
%
This study was financed in part by the Conselho Nacional de Desenvolvimento Científico e Tecnológico -- Brasil (CNPq) -- Research Project No. 440846/2023-4 and Research Fellowship No. 201221/2024-1. H.S.V. is partially supported by the Alexander von Humboldt-Stiftung/Foundation (Grant No. 1209836). Funded by the Federal Ministry of Education and Research (BMBF) and the Baden-W\"{u}rttemberg Ministry of Science as part of the Excellence Strategy of the German Federal and State Governments -- Reference No. 1.-31.3.2/0086017037.
%
%
\end{acknowledgments}
%
%

%
%
\end{document}